\documentclass[%
amsmath,amssymb,
aps,
pre,
twocolumn,
]{revtex4-2}

\usepackage{cleveref}
\crefname{algocfline}{alg.}{algs.}
\Crefname{algocfline}{Algorithm}{Algorithms}
\usepackage{graphicx}% Include figure files
\usepackage{dcolumn}% Align table columns on decimal point
\usepackage{bm}% bold math
\usepackage[ruled,vlined,linesnumbered]{algorithm2e}
\usepackage{xcolor}
\usepackage{subcaption}

\newtheorem{defn}{Definition}

\newtheorem{prop}{Proposition}

\crefname{rem}{remark}{remarks}
\crefname{defn}{definition}{definitions}

\begin{document}

\preprint{APS/001-DFT}

\title{Coarse-graining Directed Networks with Ergodic Sets\\Preserving Diffusive Dynamics}

\author{Erik Hormann}
\email{erik.hormann@ed.ac.uk}
 \affiliation{Mathematical Institute, University of Oxford}
 \affiliation{School of Mathematics, University of Edinburgh}
\author{Renaud Lambiotte}
\affiliation{Mathematical Institute, University of Oxford}

\date{\today}

\begin{abstract}
In this paper, we introduce ergodic sets, subsets of nodes of the networks that are dynamically disjoint from the rest of the network (i.e. that can never be reached or left following to the network dynamics). We connect their definition to purely structural considerations of the network and study some of their basic properties.
We study numerically the presence of such structures in a number of synthetic network models and in classes of networks from a variety of real-world applications, and we use them to present a compression algorithm that preserve the random walk diffusive dynamics of the original network.
\end{abstract}

\maketitle

%\tableofcontents

\section{\label{sec:intro}Introduction}
Directed networks have received substantial attention in recent years as they naturally arise from a number of real-world applications and  they exhibit a broader variety of phenomena than
undirected networks \cite{bang2008digraphs}. For example, transportation networks are inherently directed \cite{concas2022chained}, and so are supply chains \cite{nagurney2006supply}, food chains \cite{clemente2015formal}, and the World Wide Web \cite{broder2000graph}, just to name a few. Two of the most interesting properties of directed networks are asymmetry of the dynamics and a richer notion of connectedness \cite{rodgers2023strong}.  Asymmetry may have local consequences, such as a lack of detailed balance at stationarity \cite{nartallo2024broken}, and global ones, with the possible emergence of hierarchies \cite{mackay2020directed} and non-normal dynamics \cite{asllani2018structure} in directed networks. In addition, different forms of connectivity can be defined in networks, chiefly strong and weak connectedness. Networks that are not strongly connected are known to induce constraints on dynamics, with the existence of sources and sinks and the diffusive flow starting from the former and being absorbed in the latter, often leading to non-ergodicity. In practice, non-ergodicity makes the foundations of several network algorithms ill defined, e.g., for community detection \cite{Lambiotte2009} or node centrality \cite{gleich2015pagerank}, which can be remedied by tricks such as the use of random teleportation \cite{lambiotte2012ranking}. These asymmetric flows are also at the core of models for the visualisation of directed networks, such as the so-called bow-tie structure of the Web \cite{broder2000graph}.

In its simplest form, sources correspond to nodes with zero in-degree  and sinks to nodes with zero out-degree, which can be thought of as starting and ending points for trajectories on the directed network, thus leading to an ordering of the nodes that can be used to infer their hierarchical structure. This idea is at the core of the original formulation of trophic coherence \cite{johnson2017looplessness}.
This scenario is, however, fairly limiting, as it does not account for situations when a source or a sink is composed of multiple nodes forming, as we explain below, structures that we call forward and backward ergodic sets. This idea directly relates to the notions of absorbing states, ergodic sets, and transient states in the study of Markov chains. The main purpose of this article is to explore the notions of sources and sinks and to leverage them to introduce a dynamics-preserving coarse-graining algorithm . After introducing the mathematical concepts in Section \ref{sec:es}, we  describe algorithms to efficiently detect forward and backward ergodic sets in Section \ref{sec:detection}. Equipped with these tools, we  explore the statistical properties of sources and sinks in network models and in empirical networks. Finally, in Section \ref{sec:coarseGrain}, we investigate ways to coarse-grain an original network by collapsing nodes parts of the same source or sink and to characterise its input-output structure.

\begin{figure}[]
\includegraphics[width=0.35\textwidth]{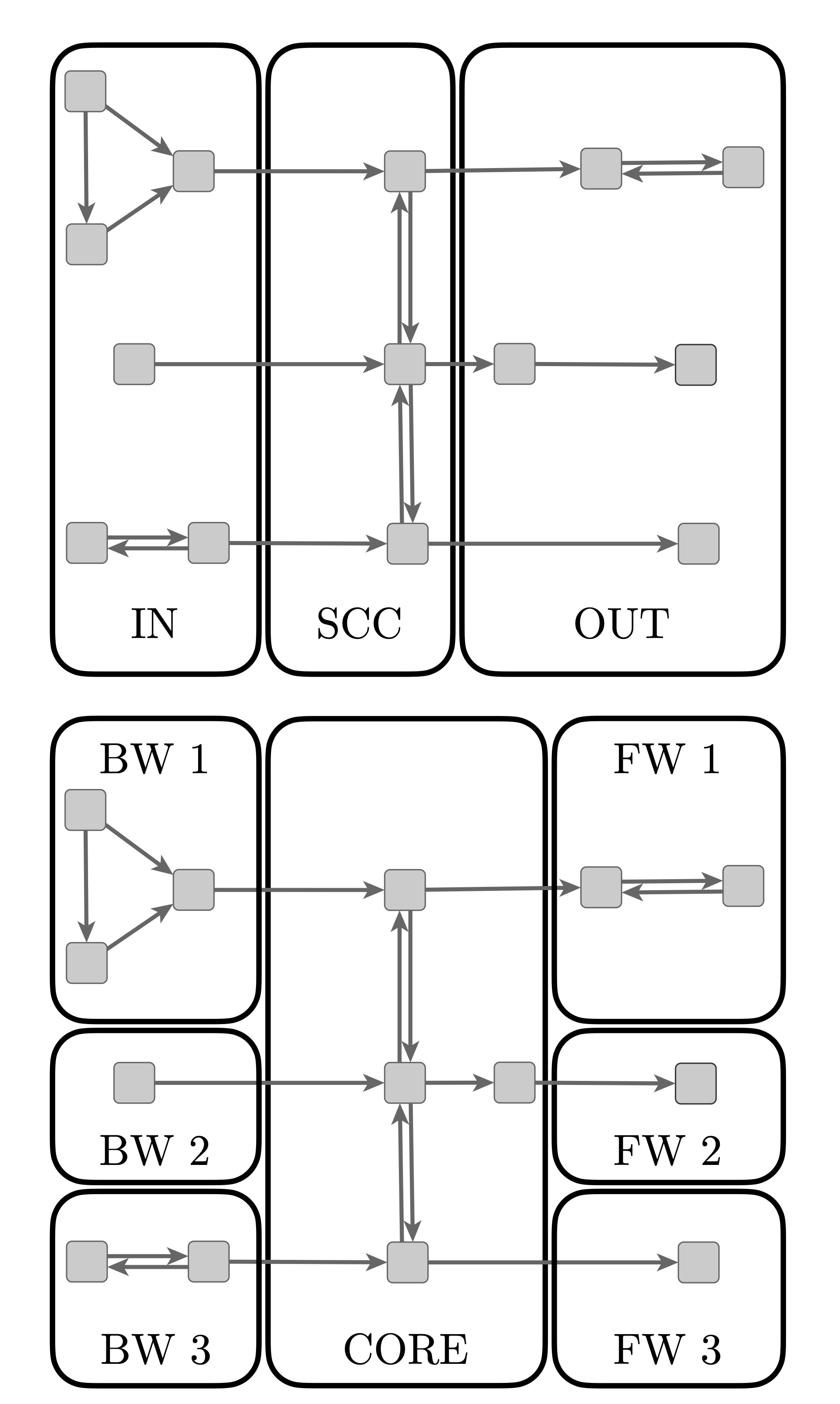}
\caption{A representation of a generic network division. \textbf{Top:} the classical \textit{bow-tie} structure \cite{broder2000graph}. \textbf{Bottom:} the division in ergodic sets. BW: backward ergodic set, FW: forward ergodic set.}
\label{fig:schema}
\end{figure}

\section{\label{sec:es}Generalised Ergodic sets}
\subsection{\label{ssec:ES}Definition and properties}
A random walk process on a network, whether directed or undirected, is equivalent to a Markov chain, with each node of the former being a state of the latter. 
Given the inter-relations between these models, but also their applications in a range of topics, a number of conflicting nomenclatures overlaps, coming from background as varied as ergodic theory, probability, graph theory, and statistical physics. Hence, we first clear up the notation and lay down some fundamental definitions to prepare for the main results that follow. 

A Markov chain on a directed network $\mathcal G$ is characterized by three constituent elements \cite{pinsky2010introduction}: the state space $V(\mathcal G)$, the stochastic transition matrix $P$, and the starting  probability distribution $\lambda(V)$.
The states of a Markov chain can be decomposed into different types \footnote{The analysis of the state space becomes more challenging in the case of infinite state spaces. In this paper, however, we  focus on finite graphs and -- therefore -- finite Markov chains.}. 
Firstly, an absorbing state of a Markov chain is a state such that once entered, the process never escapes. On a network, this corresponds to a node with no outgoing edges. 
Secondly, a node may belong to an ergodic set, which we define via the notion of ``closed communicating class''.
\begin{defn}
Let $X=(V,P,\lambda)$ be a Markov chain. One says two states $i,j \in V$ ``communicate'', and one writes $i \leftrightarrow j$ if both $\left(P_{ij}\right)^m, \left(P_{ji}\right)^n\neq 0$ for some $m,n \geq 0$.  
\end{defn}
From a network perspective, the two nodes corresponding to the states $i$ and $j$ are connected by a walk, in both directions, and they are thus strongly connected.
\begin{prop}
The property ``to communicate'' is an equivalence relation. One calls ``communicating class'' the resulting equivalence classes.
\end{prop}
By definition, each  absorbing state forms its own communicating class, but the connection between the concepts can be strengthened as follows.
\begin{defn}
Let $X\subseteq V$ be a communicating class. One says $X$ is ``closed'' if for all $i\in X$ and $j \in V \setminus X$, one has $P_{ij} = 0$.  
\end{defn}
In other words, a communicating class is \textit{closed} if no state outside the class is accessible from any state within the class. Note that the condition $P_{ij} = 0$ implies that $P^n_{ij} = 0$ for any integer $n$, and no probability \textit{leaks away} from the set at any time.  The entire communicating class now behaves like a single absorbing super-state that can never be left but in which all nodes communicate. We call a closed communicating class a \textit{forward ergodic set}, based on the property that any probability distribution defined on the nodes of a closed communicating class asymptotically converges to the same equilibrium distribution. 

Finally, a state that does not belong to an ergodic set and is thus, in particular, not an absorbing state, is called a transient state. By definition, such nodes have probability \textit{leaking away} and the probability to observe those states goes to zero when the number of iterations of the Markov chain goes to infinity. 

Ergodic sets are thus a natural way to identify sinks in a directed network. In order to identify the sources, we consider the \textit{reverse graph} \cite{essam1970some}, sometimes called the \textit{transpose graph}, that is, the directed graph obtained by $\mathcal G$ by reversing the direction of each edge.
Sources are then identified as the ergodic sets in the reverse of the graph. From now on, we will call forward ergodic sets and backward ergodic sets the ergodic sets of the original graph and its reverse respectively. More formally, they are defined as follows:

\begin{defn}\label{defn:fES}
Let $\mathcal{G}=(V,E)$ be a directed graph. We say a subgraph $X\subseteq V$ is a \textit{forward ergodic set} if $X$ is strongly connected and:
\begin{equation}\label{eq:fES}
     \forall x\in X,\,\forall v \in V\setminus X \Rightarrow (x,v) \notin E
\end{equation}
\end{defn}
 A forward ergodic set is thus a subset of the nodes that is strongly connected and has no edges leaving the set. 
 
\begin{defn}\label{defn:bES}
Let $\mathcal{G}=(V,E)$ be a directed graph. We say a subgraph $X\subseteq V$ is a \textbf{backward ergodic set} if $X$ is strongly connected and:
\begin{equation}\label{eq:bES}
    \forall x\in X,\,\forall v \in V\setminus X \Rightarrow (v,x) \notin E
\end{equation}
\end{defn}
A backward ergodic set is thus a subset of the nodes that is strongly connected and has no edges entering the set.

We call a \textit{generalised ergodic set} a subset of nodes that is either a forward or a backward ergodic set.
Generalised ergodic sets can be seen as a stricter version of strongly connected components, as they require an additional constraint on the links that enter or leave the set. Note that a generalised ergodic set is either a forward or a backward ergodic set, or -- in the case the whole graph is strongly connected -- the entire graph.

Furthermore, note that if a node belongs to a forward ergodic set and also to a backward ergodic set, then its weakly connected component is necessarily strongly connected. 
The nodes that are transient in both the original graph and its reverse are not part of a generalised ergodic set. We call their set the \textit{transient core}. 

The forward ergodic set, the backward ergodic set, and the transient core define a partition on the nodes of the network. We will use this partition to define a coarse-graining algorithm for directed network. It can be seen as an alternative of the classical bow-tie structure \cite{broder2000graph} (cfr. \cref{fig:schema}), based on the notions of sources, sinks, and their mixing via the transient core.

\subsection{\label{sec:detection}Detection algorithm}

In order to explore the structure of directed networks in terms of generalised ergodic sets, it is important to have efficient algorithms to detect them. As a first step, we therefore develop an algorithm that solves the following problem. Given a graph $\mathcal{G}$, find a list $\mathcal{X} = \{X_1,\dots,X_f\}$ of forward ergodic sets $X_i$ and a list $\mathcal{Y} = \{Y_1,\dots,Y_b\}$ of backward ergodic sets $Y_i$ in the graph. Note that $f$ and $b$, the number of forward and backward ergodic sets respectively, are not known a priori and are part of the detection problem.  

We proceed in two steps (see  \cref{alg:ESdetection} for a pseudo-code). First, we identify strongly connected components of the network, using standard algorithms \cite{tarjan1972depth}. This first step returns a list $\mathcal{C} = \{C_1,\dots,C_l\}$ in which $C_i$ are all strongly connected components of $\mathcal{G}$.
Let us focus on the detection of forward ergodic sets for the moment. From Definition 1, we know that $\mathcal{X}$ is a subset of $\mathcal{C}$. Finding the elements of $\mathcal{X}$ thus requires us to check for every element $C_i$ of $\mathcal{C}$ whether its nodes have edges leaving $C_i$. In practice, we implement this second step as follows. 
Each element $X \in \mathcal{C}$ is a subgraph of the original graph, to which one can make correspond an induced subgraph $X_I$, that is a subgraph built from the nodes of $X$ and all of their connections - not only the connections between elements of $X$. It is straightforward to show that 
\begin{equation}
\sum_{v\in X} d_{out}(v) - \sum_{v\in X_I} d_{out}(v) = \mbox{\# edges from }X \mbox{ to }\mathcal{G}\setminus X,  \notag
\end{equation}
where $d_{out}(v)$ is the out-degree of node $v$ in the corresponding subgraph.
For each connected component $X$, we therefore compute the quantity $\sum_{v\in X} d_{out}(v) - \sum_{v\in X_I} d_{out}(v)$.
If it is equal to zero, $X$ is identified as a forward ergodic set, and it is added to $\mathcal{X}$. 
For backwards ergodic sets, we perform the same operation, but this time focusing on incoming edges, hence computing $\sum_{v\in X} d_{in}(v) - \sum_{v\in X_I} d_{in}(v)$, and adding a backward ergodic set to $\mathcal{Y}$ if this difference is equal to zero. Note that a set can in principle be both a forward and a backward ergodic set. If the set is neither of those, the nodes of $X$ are added to the transient core.

\begin{algorithm}[H]
\label{alg:ESdetection}
\SetKwInOut{Input}{Input}
\SetKwInOut{Output}{Output}
\Input{$\mathcal{G}(V,E), V \supset X$ strongly connected}
\Output{Yes/No}
\DontPrintSemicolon
\caption{Detecting a Forward Ergodic Set}
\SetKwFunction{IsFwdErgodicSet}{IsFwdErgodicSet}
\SetKwProg{Fn}{Function}{}{}
\BlankLine
\Fn{\IsFwdErgodicSet{$X$}}{
    $X_I \gets \mathcal{G}[X]$
    $\Delta_X \gets 0$\;
    \For{$x \in X$} {
        $\Delta_X  \gets \Delta_X + d_X(x) - d_{X_I} (x)$\;
    }
    \If{$\Delta_X = 0$}{
        \KwRet True\;
    }\Else{
        \KwRet False\;
    }
}
\end{algorithm}

\section{\label{sec:coarseGrain}Corse-graining with ergodic sets}

The algorithm described in the previous section takes a directed graph $\mathcal G$ and partitions its nodes into sets  that have the same properties in terms of reachability with respect to a random walk. That is, all the nodes in a forward ergodic set have the same probability of being asymptotically attained from any other node in the directed network. Our purpose is to use these sets in order to coarse-grain  dynamical processes, with the purpose of obtaining a faithful representation of a random walk on a compressed version of the original network \footnote{There are multiple motivations for this, from the theoretical foundations of Markov chains to the use of random walks in network algorithms.}. From now on, we are assuming weak connectedness: if $\mathcal{G}$ is not weakly connected, we can just treat each weakly connected component separately; if $\mathcal{G}$ is strongly connected, the whole graph is a forward and backward ergodic set, and the decomposition is trivial.

Our method, which we call \textit{ergodic set compression algorithm} (ESCA), is divided into two steps. 
It takes as an input the generalised ergodic sets and the transient core identified in the previous section.
The first step assigns weighted connections between the generalised sets and the nodes in the core in order to preserve the random walk dynamics. 
The second step considers the transient core as a ``black box" that mixes probability between the backward and the forward ergodic sets. The resulting representation of the core as a rectangular matrix can then be used to compress the transient core into a small number of latent states, thereby allowing to uncover different modes through which backward ergodic sets communicate with forward ergodic sets.
Both steps preserve the network flows by construction, but only the first is lossless. The first step preserve the flows and the transition times between nodes. In contrast, the second step provides an approximation of the flows in the long time limit $t \rightarrow \infty$.

\subsection*{Step 1: Ergodic sets compression.}
The fundamental idea is to coarse-grain the strongly-connected ergodic sets into single nodes and modify the network edges in order to preserve the marginal transition probability between generalised ergodic sets and the transient core, and vice-versa. The compression rules are generated separately for forward ergodic sets and backward ergodic sets.  For forward ergodic sets, we equate the probabilistic 1-step in-flow from all the nodes that are the origin of any edge ending in one node of the (forward) ergodic set. For backward ergodic sets, we simply copy the outflow from any node that has been clustered into the input. We now provide the complete rewiring rules.

Let $X$ be a \textit{forward} ergodic set. We replace $X$ by a single node $v_X$, which inherits incoming links from all the nodes $w \in V \setminus X$ for which -- before the replacement -- there exists $v\in X$ such that $( w,v) \in E(\mathcal{G})$.  A weight is assigned to each edge as usually done when coarse-graining networks \cite{blondel2008fast}: from a given $w$, it is equal to the number of links from $w$ to the nodes in $X$ (in the case of weighted networks, it would be the sum of the weights of links to nodes in $X$).  The following pseudo-code illustrate the algorithm step-by-step.

\begin{algorithm}[H] 
\label{alg:coarseFW}
\SetKwInOut{Input}{Input}
\SetKwInOut{Output}{Output}
\Input{$\mathcal{G}(V,E,\omega), X \subset V$}
\Output{\{$\omega(y,X) \; \forall y \in V \setminus X$\}}
\DontPrintSemicolon
\caption{Coarse-Graining a Forward Ergodic Set}
\SetKwFunction{FWeight}{Weights}
\SetKwProg{Fn}{Function}{}{}
\BlankLine
\Fn{\FWeight{$X$}}{
    \For{$y \in V \setminus X$} {
        $\omega(y, X) \gets 0$\;
        \For{$x \in X$}{
            \If{$(y,x) \in E$}{
                $\omega(y,X) \gets \omega(y,X) + \omega(x,y)$
            }
        }
    }
    \KwRet {$\omega(X, y) \; \forall y \in V \setminus X$}\;
}
\end{algorithm}

\begin{figure}[ht]
  \includegraphics[width=.75\linewidth]{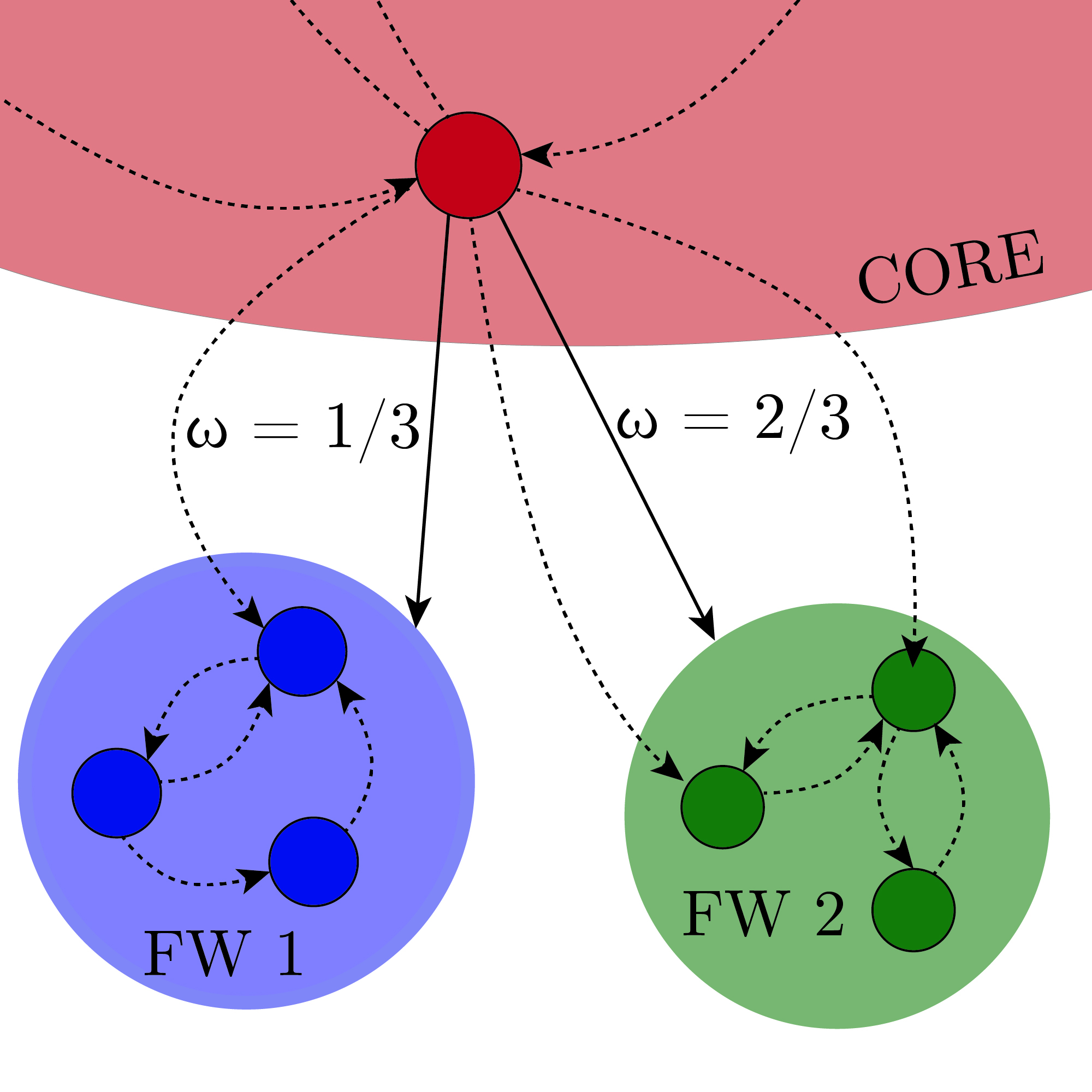}
  \label{fig:FWES}
  \centering
\caption{\label{fig:epsart} Example of \cref{alg:coarseFW} applied to a forward ergodic set. The dashed edges to nodes of the forward ergodic sets are replaced by the plain edges to the ergodic sets. The latter are weighted such as to preserve the flows of random walkers. }
\end{figure}

The problem is very similar, yet slightly more subtle in the case of backward ergodic sets.
Let $Y$  be a \textit{backward} ergodic set. 
As before, we collapse its vertices and define a new meta-node  $n_Y$. 
This node inherits the connections of the nodes that were part of the corresponding ergodic set. The weights are then assigned such as to conserve the  flow of the random walk. Let us consider a node 
$v \in V\setminus Y$. The flow of probability to that node is not simply the number of edges from nodes in $Y$. As nodes in $Y$ are dynamically independent from those in the rest of the graph, one needs to specify their density of walkers. In addition, one has to consider the fact that the probability leaving one of those nodes is distributed along its outgoing edges. 
At a given time $t$, the flow is the product of the probability $p(y,t)$ of a random walker being on node $y \in Y$ and the transition probability $p(y \rightarrow v)$ toward a node $v$ outside the set. 
Given some initial condition for the nodes of a backward ergodic set, and as the probability progressively leaks out of the set, the probability to be on each node evolves in time. Importantly, this means that the flows may be biased toward certain nodes of the core at certain times, and to other nodes at other times.
The probability to observe flows between the set and node $v$ are obtained by summing over their nodes, and thus also depend on time in general. Our purpose is to build a time-independent coarse-graining of the network, hence we need to make a sensible choice for a time-independent version of $p(y,t)$. Possibilities  include a uniform distribution or the stationary distribution of the random walk defined on the ergodic set (without leakage). 
The latter choice is an adiabatic approximation in which the dynamics thermalise on the ergodic set before probability leaks to the external world. Another possibility would be to consider the dominant eigenvector of the transition matrix associated to the induced subgraph of the ergodic set (thus with leakage), which would now characterise how, asymptotically, the probability approaches zero on each  node of the set. For the sake of simplicity, and because it is usually a first-order approximation of these asymptotic regimes \cite{lambiotte2012ranking}, we simply consider here $p(y,t)$ to be proportional to the in-degree of node $y$.

For this choice, 
the weight from set $Y$ to node $v$ takes the form
\begin{align}
    \omega_{Yv} &= \sum_{y\in Y} \frac{\mathbb{I}_E(y,v)}{d_{out}(y)}p(y) =\\
    &= \frac{1}{\sum_{y\in Y}d_{in}(y)}\sum_{y \in Y}\frac{d_{in}(y)}{d_{out}(y)}\mathbb{I}_E(y,v)
\end{align}
where the indicator function $\mathbb{I}_E(y,v)$is defined as:
$$\mathbb{I}_E(y,v) =\begin{cases}
1 \mbox{ if } (y,v) \in E\\
0 \mbox{ otherwise}
\end{cases}$$ 
A pseudo-code is given in \cref{alg:coarseBtW}.
The aggregated weights depend only on the original adjacency matrix. Furthermore, they are calculated only from local properties of the ergodic set (namely, the adjacent edges).
%, as can be seen using the adjacency matrix (see Fig.\ref{fig:ES_AdjMatrix}).

Locality  implies that the detection and the coarse-graining algorithms are fast. The Tarjan algorithm \cite{tarjan1972depth} used to detect strongly connected components is $O(V+E)$, while the adjacency matrix check is $O(M_1N)$, in which $M_1$ is the size of the largest ergodic set. Crucially, in most applications the largest strongly connected is within the graph core, for which the algorithm will quickly find an outgoing and incoming edge,
%-- a nonzero element in the blue and red areas of \cref{fig:ES_AdjMatrix}, 
after which the algorithm will stop.
If we assume the graph to have a giant strongly connected component, therefore, the detection algorithm in \cref{alg:ESdetection} will instead run in $O(M_2N)\sim O(N$).

Note that, while the detection of backward ergodic sets could be run using the same algorithm as for the forward ones on the inverse network, it is not possible to do so for the coarse-graining. This would fail to account for the difference in the site probability $p(y)$ for $y\in Y$. In other words, while detection is perfectly symmetric, compression is not.
%This is due to the fact that, inside a backwards ergodic set, different nodes will have different probability $p(y,t)$ for the location of the thermalised random walk.

\begin{algorithm}[ht]
\label{alg:coarseBtW}
\SetKwInOut{Input}{Input}
\SetKwInOut{Output}{Output}
\Input{$\mathcal{G}(V,E,\omega), X \subset V$}
\Output{\{$\omega(X,y) \; \forall y \in V \setminus X$\}}
\DontPrintSemicolon
\caption{Coarse-Graining a Backward Ergodic Set}
\SetKwFunction{FWeight}{Weights}
\SetKwProg{Fn}{Function}{}{}
\BlankLine
\Fn{\FWeight{$X$}}{
    \For{$y \in V \setminus X$} {
        $\omega(X,y) \gets 0$\;
        \For{$x \in X$}{
            \If{$(x,y) \in E$}{
                $p(x) \gets d_{in}(x)/\sum_{x'\in X}d_{in}(x') $ \\
                $\omega(X,y) \gets \omega(X,y) + p(x)\omega(x,y)$
            }
        }
    }
    \KwRet {$\omega(X, y) \; \forall y \in V \setminus X$}\;
}
\end{algorithm}

%\begin{figure}[]
%\includegraphics[width=.40\textwidth]{./figures/ESMatrix.png}
%\caption{Given the graph adjacency matrix $A$ and the ergodic set $X$, green in the figure, we only need to check the matrix elements in blue (for a forward ergodic set) or red (for backward ergodic set} 
%\label{fig:ES_AdjMatrix}
%\end{figure} 

\begin{figure}[]
  \includegraphics[width=.75\linewidth]{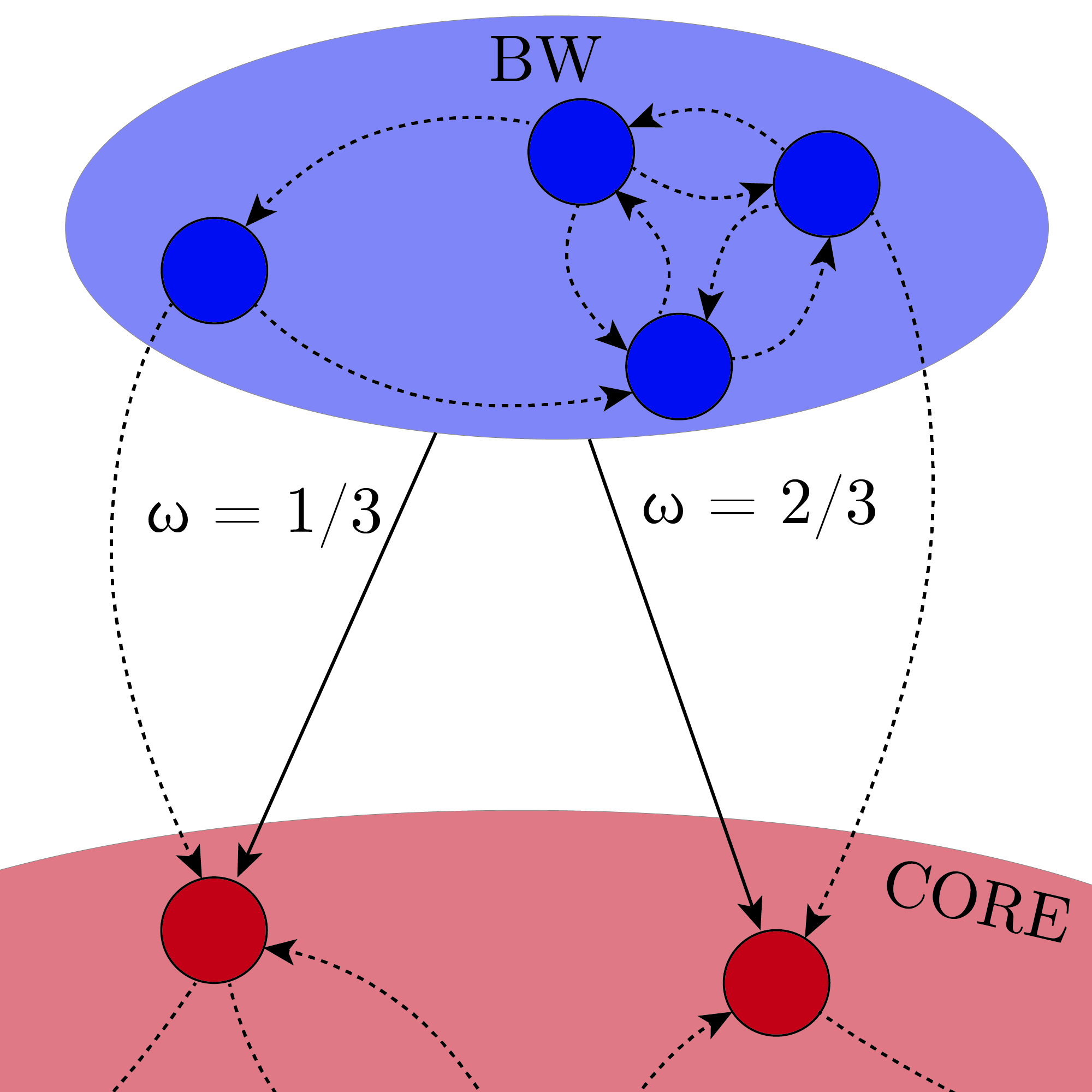}
\label{fig:BWES}\caption{\label{fig:inputs} Example of \cref{alg:coarseBtW} applied to a backward ergodic set. Dashed edges from nodes in backward ergodic sets are replaced by plain edges from the ergodic sets, with weights chosen to preserve flows of walkers. }
\end{figure}

%\begin{figure}[ht]
%\begin{subfigure}{.24\textwidth}
%  \centering
%  \includegraphics[width=.9\linewidth]{./figures/input_2.jpg}
%  \label{fig:input2}
%\end{subfigure}%
%\begin{subfigure}{.25\textwidth}
%  \centering
%  \includegraphics[width=.9\linewidth]{./figures/input_3.jpg}
%  \label{fig:input3}
%\end{subfigure}
%\caption{\label{fig:inputs} Example of \cref{alg:coarseBW} applied to a backward ergodic set. Grey edges and nodes with grey borders disappear during the rewiring; black edges and elements with black borders are introduced by the rewiring.}
%\end{figure}

\subsection*{Step 2: Core compression.}
At the end of the first step, we have replaced the original network with a  compressed network where each ergodic set, forward or backward, is replaced by a single node. 
By construction, these nodes are sinks (having zero out-degree) and sources (having zero in-degree) respectively. 
We denote by $N_{bw}$, $N_{fw}$ and $N_{core}$  the number of sources (i.e. coarse-grained backwards ergodic sets), sinks (i.e. coarse-grained forwards ergodic sets) nodes in the core respectively.
At this stage, we have retained all the nodes that are part of the transient core, hence faithfully describing how the random walk moves from the sources to the sinks. 
Each of those three network component is associated with a corresponding transition probability for the random walk on the network. Hence, the dynamic on the network can be decomposed into three regimes: the transition from the backward ergodic sets to the core, the mixing in the core, and the transition towards the forward ergodic sets. These three regimes take place one after the other and the transition from one to the next is governed by the hitting times of the network core and the forward ergodic sets. Because node hitting times are stopping times in the sense of Markov chains \cite{norris1998markov}, the overall random walk transition probability can be factorised into three distinct probabilities.

Inspired by this result, we consider the asymptotic behaviour of the random walk and we compress the core as a black box connecting backward and forward ergodic sets, respectively interpreted as sources and sinks. 
Starting from the compressed network obtained in Step 1, we  consider a rectangular matrix capturing the probability to finish at a certain sink when starting on a certain source, and then compress it by uncovering the dominant mixing patterns in the core.
The \textit{core mixing matrix} is defined as:

\begin{equation}
    B_{ij} = \lim_{t\rightarrow \infty} p(y_j,t|x_i,0),
\end{equation}
that is, the probability to be absorbed by one of the sinks $y_i$ when starting on a source $x_i$.
By construction, $\sum_j B_{ij}=1$. 

A natural way to reduce the dimensionality of the system even further and to uncover the dominant patterns in the mixing matrix is to perform a singular values decomposition (SVD) \cite{martin2012extraordinary} as follows:
\begin{equation} 
\label{eq:decomposition}
    B = M_{bw}\,C\,M_{fw}.
\end{equation}
where $M_{bw} \in \mathbb{R}^{N_{bw} \times k}$ and   $M_{fw} \in \mathbb{R}^{k \times N_{fw}}$ encode the positions of the sources and the sinks in a $k$-dimensional latent space, and where the diagonal matrix $C \in \mathbb{R}^{k\times k}$ encodes the importance of each dimension in that space. 

In doing this, from an original network of $N$ nodes, we have first constructed a compressed network made of $N_1=N_{bw}+N_{fw}+N_{core} \leq N$ nodes, and then, as a second step, a more compressed system composed of $N_2=N_{bw}+k+N_{fw}$. 
Note that the factorisation (\ref{eq:decomposition}) can be performed exactly, hence leading to optimal, lossless compression, in which case $k \leq \min(N_{bw},N_{fw})$ is equal to the rank of the core mixing matrix, but that it can also serve as a basis for a further, lossy compression that can be obtained by keeping only a subset  of dominant dimensions.

\section{\label{sec:NA} Numerical experiments} 
In order to assess the algorithm performance, we apply it to a number of different networks, both synthetic random graphs and real-world networks taken from various areas of applications. We test the two compression steps separately, so that we are able to estimate the impact of each part of the algorithm. The main metric we use to score the algorithm are the \emph{compression factors} $C_1$ and $C_2$,  which represent the relative reduction in the number of nodes in the network after the first and the second step of the algorithm, respectively. More explicitly, if $N$ is the number of nodes in the original network, $N_1$ and $N_2$ are the number of nodes in the network after the first and second coarse-graining step, as defined above, and the two compression factors are defined as follows:
\begin{equation}
    C_1 = 1-\frac{N_1}{N} \qquad C_2 = 1-\frac{N_2}{N}.
\end{equation}
The compression factors are thus $0\leq C_1\leq C_2 \leq 1$ by definition, and higher compression factors correspond to a higher compression power of the algorithm.

\subsection{\label{subsec:sinthetic} Generalised ergodic sets in random networks}

In random graphs, the emergence of generalised ergodic sets is a random process directly related to percolation theory \cite{callaway2000network} and the formation of a giant strongly connected component \cite{dorogovtsev2001giant}. Both phenomena have been investigated in detail and are know analytically \cite{cohen2021percolation}, making them a baseline \emph{ground truth} to compare against.

In the undirected case, and for the simplest case of the Erd\H{o}s-R\'enyi random graph $ER(N,p)$, the critical probability threshold has been calculated exactly in the limit of infinitely large size, and the critical threshold sits at $p_C = 1/N$ \cite{albert2002statistical}. Below that point, the graph is composed of a multitude of small, tree-like disconnected subgraphs. For $p$ above the critical value, the giant connected component of size $O(N)$ appear in the graph.

\begin{figure}[]
\begin{subfigure}{0.45\textwidth}
\includegraphics[width=\textwidth]{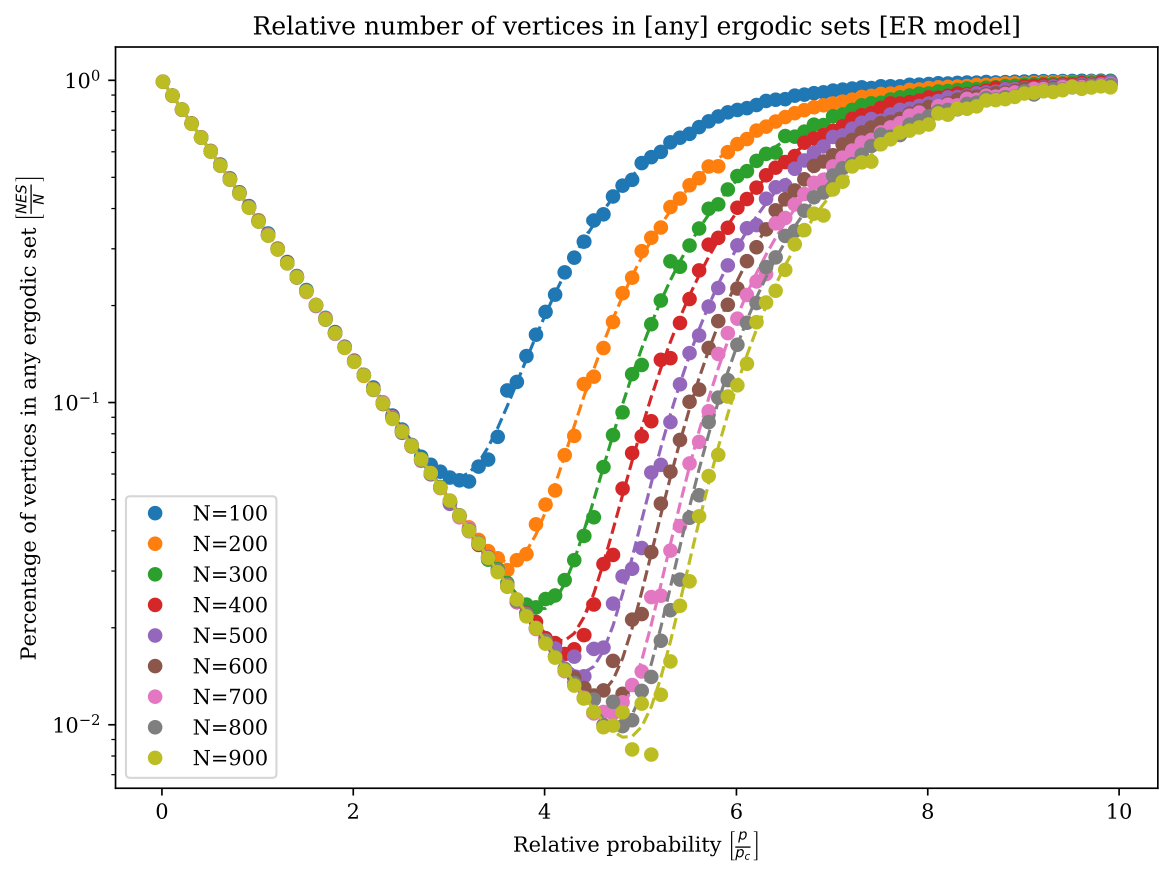}
%\caption{Fraction of nodes of an Erd\H{o}s-R\'enyi random graph belonging to any ergodic set, for different connection probabilities $p$ and network sizes $N$.}
\end{subfigure}
\begin{subfigure}{0.45\textwidth}
\includegraphics[width=\textwidth]{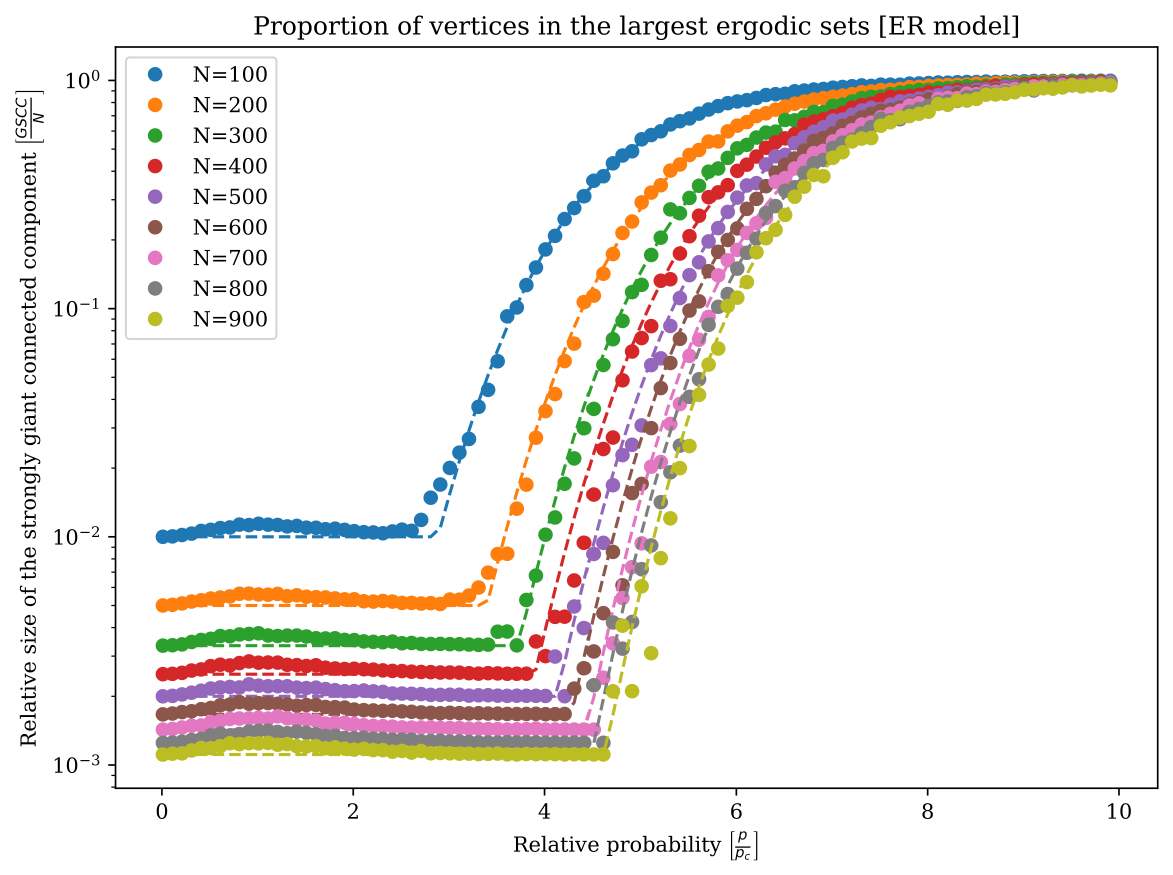}
%\caption{Size of the largest ergodic set in an Erd\H{o}s-R\'enyi random graph, for different connection probabilities $p$ and network sizes $N$.}
\end{subfigure}
\caption{Fraction of nodes of an Erd\H{o}s-R\'enyi random graph belonging to any ergodic set (above) and the largest ergodic set (below), for different connection probabilities $p$ and network sizes $N$.}
\label{fig:ER}
\end{figure}

The problem become slightly more complicated in the case of directed networks, with the emergence of weak and strong giant connected components at two distinct critical thresholds \cite{dorogovtsev2001giant}. For ergodic sets, one can distinguish three regimes:  for $p \ll p_C$, the network is essentially composed of a point cloud of isolated nodes: as each isolated node is by definition an ergodic set, the number of nodes belonging to a forward or backward ergodic set is close to $N$. In the other limit,  for $p \gg p_C$, the network is strongly connected, so the giant strongly connected component spans the entire graph. In this second case, the system is composed of one giant ergodic set and therefore the number of nodes belonging to a forward or backward ergodic set is, once again, close to $N$. 
This behaviour is illustrated in \cref{fig:ER}, where we average results over 1000 instances of   Erd\H{o}s-R\'enyi graphs, for a range of values of  number of nodes $N$ and connection probability $p$.
Interesting behaviours emerge between these extreme values. As $p$ increases from very small values, nodes start to connect, forming many small weakly connected components, such that more and more nodes become part of the transient core, until the formation of strongly connected components, and a decrease of the core until all nodes are part of the giant strongly connected component.

This result emphasises that generalised ergodic sets are out-of-equilibrium phenomena: while the number of nodes across different instances of the random graphs stabilise to a threshold, the actual nodes that belong to those sets vary from one random graph to the next. As such, we do no investigate the compression from the second step of the algorithm, as there is no intrinsic directedness in the graph, on top of the one induced by noise.

\subsection{\label{subsec:realworld} Generalised ergodic sets in real world networks}
We test our method on a range of real-world, directed networks taken from different applications.  We have chosen a mix of networks from biology, information technology, and transportation. In our selection, biological networks represent protein-protein interaction (PPI) networks, which have been extensively studied \cite{rao2014protein,athanasios2017protein} and whose characteristics have been thoroughly explored. They represent large, densely connected networks with a high average degree. The transportation networks used are airlines networks \cite{burghouwt2005temporal}, which are known to possess a few, highly connected hubs that facilitate the transfer of passengers. Finally, for technological networks, we have chosen a subset of the  World Wide Web, specifically the links between the institutional websites of the state agencies of each of the 50 states in the United States of America \cite{kosack2018functional}. Web networks tend to possess fat-tailed degree distributions, well approximated by a power-law.

Together, these three classes of networks include a wide range of characteristics and degree distributions, which is the only information needed for the detection algorithm. Thus, having a large variety of degree distributions is important to test both the prevalence of ergodic sets and the efficiency of the detection algorithm.

\begin{figure}[]
\includegraphics[width=0.48\textwidth]{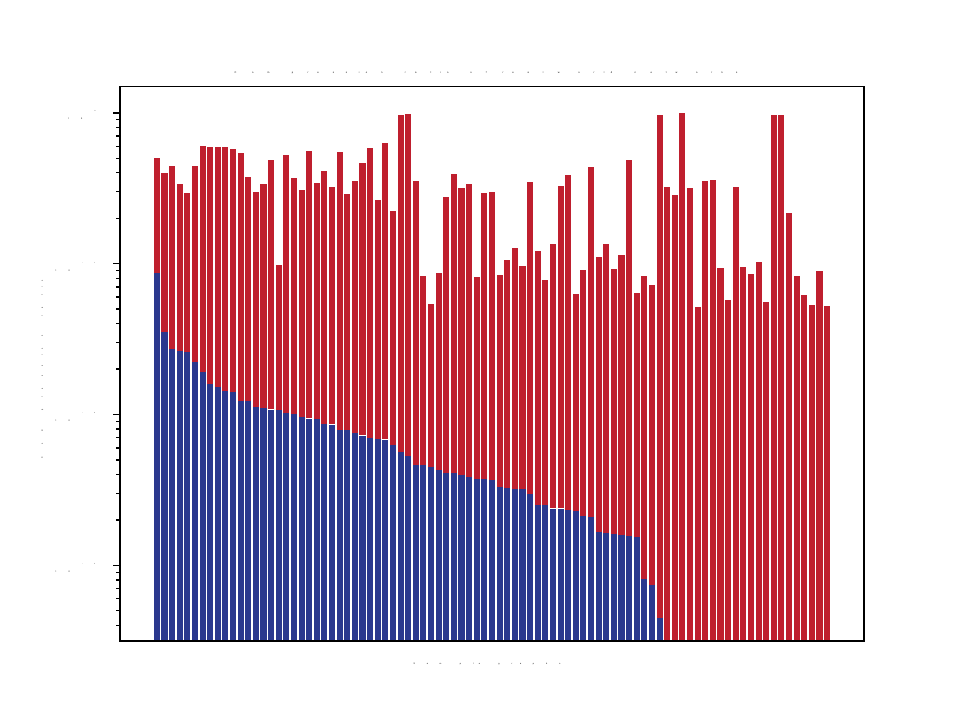}% Here is how to import EPS art
\caption{\label{fig:histC1} Compression factor $C_1$ for the first step of the ESCA. The blue data is the compression factor $C_1$, the red data is the total compression factor $C_2$.}
\end{figure}

The results are reported in \cref{fig:histC1} for both the compression rate $C_1$ after the first step of the algorithm and for the final compression coefficient $C_2$, as defined at the beginning of Section \ref{sec:NA}.

The results show that the first step of the ESCA really is a structural process, whose main aim is to simplify the graph structure and preparatory for the second step, which performs the bulk of the compression. For reference, on the dataset that we have analysed, the average compression after the first step hovers around 1\%, with the final compression in the 50\%.

For more data about the results of our numerical simulations, please refer to  \cref{tab:data} in Appendix A.

\subsection{Core-periphery structure}
%\subsection{Core mixing strength}
The second step of the algorithm substitute the graph core into set of disconnected nodes. While the long-time asymptotic flows are conserved by construction, it is natural to ask how much information is destroyed in the process. In this section, we try to quantify this by measuring the structure of the network core before rewiring. To do so, we study the dynamical distance of the core before the second ESCA step is performed with respect to a random graph.

We measure the mixing power of the network core using the simmetrised version of the core mixing matrix $B$: $\mathcal{B}=B^T B$.
In particular, we study the spectral gap of its Laplacian:
\begin{equation}
    L\left(\mathcal{B}\right)_{ij} = \sum_k B_{ik} \delta_{ij} - B{ij} 
\end{equation}
 We compare its spectrum $\sigma(L)$ to the spectrum of a graph in which the core has been rewired using a directed configuration model, which preserve the core degree distribution by construction. The rewiring has only effected the edges in the CORE part of the network (see \cref{fig:schema}), and the rewiring has been done using a directed configuration model in which self-loops were not allowed. Edges crossing a region boundary in \cref{fig:schema} were not rewired. Our findings indicate differences that are well over the threshold of statistical significance.

\begin{figure}[ht]
\begin{subfigure}{.245\textwidth}
  \centering
  \includegraphics[width=.9\linewidth]{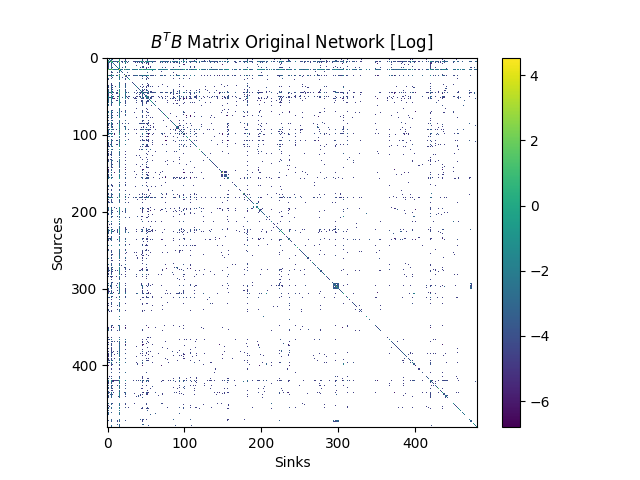}
  \label{fig:california}
\end{subfigure}%
\begin{subfigure}{.245\textwidth}
  \centering
  \includegraphics[width=.9\linewidth]{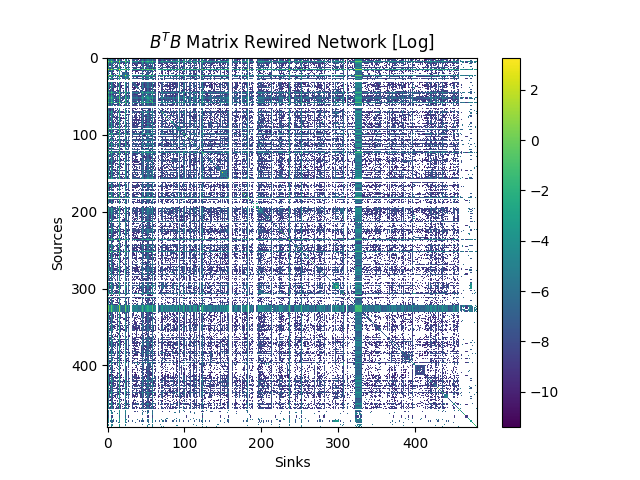}
  \label{fig:californiaR}
\end{subfigure}
\begin{subfigure}{.245\textwidth}
  \centering
  \includegraphics[width=.9\linewidth]{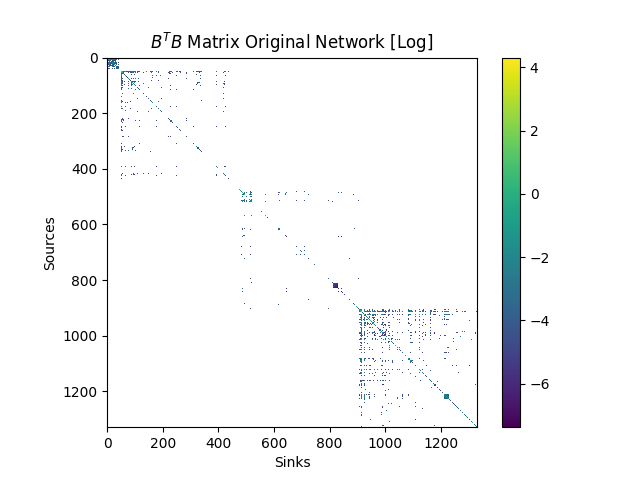}
  \label{fig:oregon}
\end{subfigure}%
\begin{subfigure}{.245\textwidth}
  \centering
  \includegraphics[width=.9\linewidth]{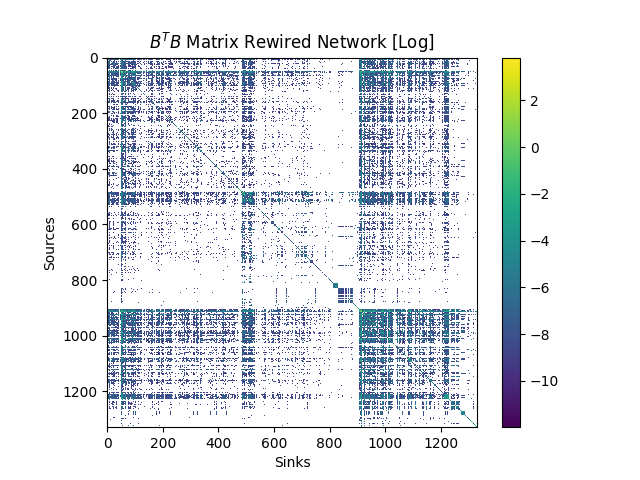}
  \label{fig:oregonR}
\end{subfigure}
\begin{subfigure}{.245\textwidth}
  \centering
  \includegraphics[width=.9\linewidth]{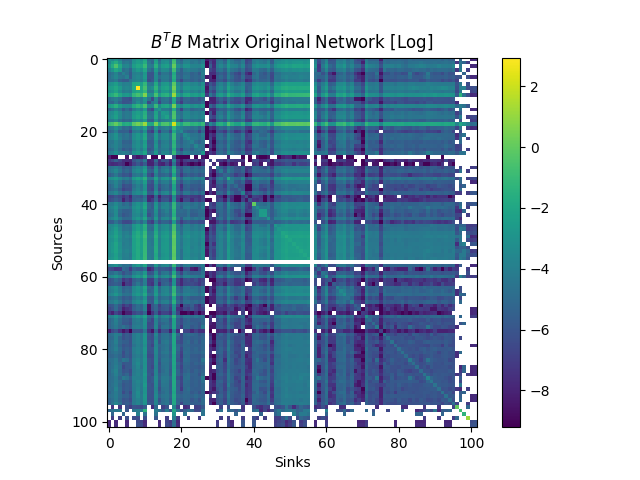}
  \label{fig:utah}
\end{subfigure}%
\begin{subfigure}{.245\textwidth}
  \centering
  \includegraphics[width=.9\linewidth]{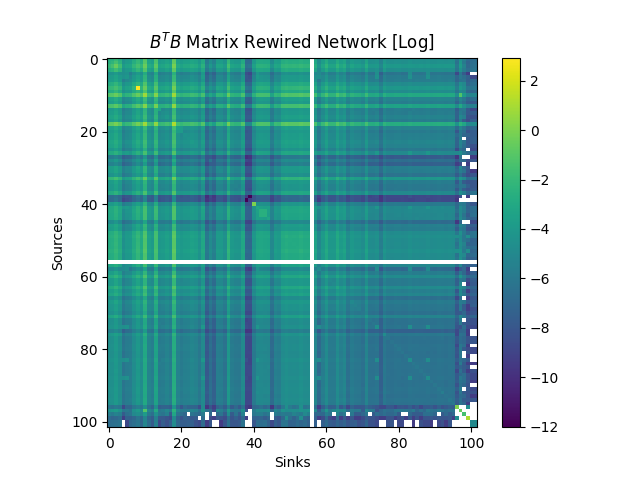}
  \label{fig:utahR}
\end{subfigure}
\caption{Plots of the symmetrised input-output connection matrix $\mathcal{B}=B^TB$. To the left the plots for the actual networks, to the right the plots for the same networks with rewired cores. By row, from top to bottom, the networks represent the links between the websites of the states agencies of California, Oregon, and Utah.}
\label{fig:BTBplots}
\end{figure}

In \cref{fig:BTBplots} the $\mathcal{B}$ results are shown for three sample networks and their rewired counterparts: directed networks representing the links between the institutional websites of the state agencies  of California, Oregon, and Utah.

Rewiring the core has invariably determined an increase in the mixing capability of the core, showing that the network cores are different from their rewired counterparts to a statistically significant degree. The average number of zero eigenvalues of $L\left(\mathcal{B}\right)$ has decreased from 100.0 (before rewiring) to 13.7 (after rewiring), with standard deviation 39.1 and 6.7 respectively. This is a statistically significant reduction: using a t-distribution with 51 degrees of freedom and the t-statistic of 15.24, we find a p-value equal to 0, up to the limits of numerical precision.

This result reinforces the idea that the second step of the compression algorithm is well posed, as there is residual excess structure in the specific connection of the core. To further corroborate this result, we study the correlation between the decrease in the number of zeroes in the spectrum of the graph Laplacian before and after the core rewiring and the compression factor $C_2$. We find a correlation coefficient $\gamma = -0.75$, indicating strong anti-correlation between the two quantities. In other words, graphs whose core is further away from a random graph are more difficult to compress.

\section{Discussion}
In this paper, we have introduced an algorithm to compress directed networks using ergodic sets, which preserves the random walk dynamics. We started by introducing the concept of generalised ergodic sets, a generalization of sinks and sources in directed networks. We used such structures to perform a first compression step that reduces the complexity of the original network into single nodes structures whose hitting times factorise the dynamics. We then used such dynamical factorization to design a second compression step based on the singular values decomposition of the matrix representing the long-time limit of the probability distribution. 

There other methods, mostly community-based, that are able to coarse-grain the network and are faster and more efficient than our approach. Some can even generate a hierarchy of progressively more coarse-grained networks. This is the case, for example, of the Louvain \cite{blondel2008fast} and the Leiden \cite{traag2019louvain} algorithm. However, the power of the method  proposed in this paper lies in the dynamical properties of the resulting networks. In other coarse-graining algorithms, there is no guarantee for the dynamics after the coarse-graining to be preserved after compression. In other words,  existing methods do not take into account the resulting dynamics when performing the network compression. Our method, on the other hand, guarantees the coarse-grained result to have the same dynamical properties of the original system. This, combined with its light computational cost and quick runtime, makes it an excellent first step in any dynamic-based investigation of directed networks.

Finally, we have applied our algorithm to a wide class of synthetic and real-world networks. We have found good agreement with the percolation threshold in synthetic networks and excellent compression power for real networks from a variety of different domains.

\begin{acknowledgements}
R.L. acknowledges support from the EPSRC grants EP/V013068/1, EP/V03474X/1 and EP/Y028872/1.

\end{acknowledgements}

\bibliographystyle{numeric}
\bibliography{ErgodicSet}% Produces the bibliography via BibTeX.

\begin{thebibliography}{10}
\expandafter\ifx\csname url\endcsname\relax
  \def\url#1{\texttt{#1}}\fi
\expandafter\ifx\csname urlprefix\endcsname\relax\def\urlprefix{URL }\fi

\bibitem{bang2008digraphs}
J.~Bang-Jensen and G.~Z. Gutin, \textit{Digraphs: theory, algorithms and applications}. Springer Science \& Business Media (2008).

\bibitem{concas2022chained}
A.~Concas, C.~Fenu, L.~Reichel, G.~Rodriguez, and Y.~Zhang, Chained structure of directed graphs with applications to social and transportation networks. \textit{Applied Network Science} \textbf{7}(1), 64 (2022).

\bibitem{nagurney2006supply}
A.~Nagurney, \textit{Supply chain network economics: dynamics of prices, flows and profits}. Edward Elgar Publishing (2006).

\bibitem{clemente2015formal}
F.~Clemente, P.~A. Nasuelli, and R.~Baggio, Formal network analysis of a food supply chain system: a case study for the italian agro-food chains. \textit{Journal of Agricultural Informatics} \textbf{6}(4) (2015).

\bibitem{broder2000graph}
A.~Broder, R.~Kumar, F.~Maghoul, P.~Raghavan, S.~Rajagopalan, R.~Stata, A.~Tomkins, and J.~Wiener, Graph structure in the web. \textit{Computer networks} \textbf{33}(1-6), 309--320 (2000).

\bibitem{rodgers2023strong}
N.~Rodgers, P.~Ti{\v{n}}o, and S.~Johnson, Strong connectivity in real directed networks. \textit{Proceedings of the National Academy of Sciences} \textbf{120}(12), e2215752120 (2023).

\bibitem{nartallo2024broken}
R.~Nartallo-Kaluarachchi, M.~Asllani, A.~Goriely, and R.~Lambiotte, Broken detailed balance and entropy production in directed networks. \textit{arXiv preprint arXiv:2402.19157}  (2024).

\bibitem{mackay2020directed}
R.~S. MacKay, S.~Johnson, and B.~Sansom, How directed is a directed network? \textit{Royal Society open science} \textbf{7}(9), 201138 (2020).

\bibitem{asllani2018structure}
M.~Asllani, R.~Lambiotte, and T.~Carletti, Structure and dynamical behavior of non-normal networks. \textit{Science advances} \textbf{4}(12), eaau9403 (2018).

\bibitem{Lambiotte2009}
R.~Lambiotte, J.-C. Delvenne, and M.~Barahona, {L}aplacian {D}ynamics and {M}ultiscale {M}odular {S}tructure in {N}etworks. \textit{arxiv:0812.1770} pp. 1--29 (2009).

\bibitem{gleich2015pagerank}
D.~F. Gleich, Pagerank beyond the web. \textit{siam REVIEW} \textbf{57}(3), 321--363 (2015).

\bibitem{lambiotte2012ranking}
R.~Lambiotte and M.~Rosvall, Ranking and clustering of nodes in networks with smart teleportation. \textit{Physical Review E—Statistical, Nonlinear, and Soft Matter Physics} \textbf{85}(5), 056107 (2012).

\bibitem{johnson2017looplessness}
S.~Johnson and N.~S. Jones, Looplessness in networks is linked to trophic coherence. \textit{Proceedings of the National Academy of Sciences} \textbf{114}(22), 5618--5623 (2017).

\bibitem{pinsky2010introduction}
M.~Pinsky and S.~Karlin, \textit{An introduction to stochastic modeling}. Academic press (2010).

\bibitem{Note1}
The analysis of the state space becomes more challenging in the case of infinite state spaces. In this paper, however, we focus on finite graphs and -- therefore -- finite Markov chains.

\bibitem{essam1970some}
J.~W. Essam and M.~E. Fisher, Some basic definitions in graph theory. \textit{Reviews of Modern Physics} \textbf{42}(2), 271 (1970).

\bibitem{tarjan1972depth}
R.~Tarjan, Depth-first search and linear graph algorithms. \textit{SIAM journal on computing} \textbf{1}(2), 146--160 (1972).

\bibitem{Note2}
There are multiple motivations for this, from the theoretical foundations of Markov chains to the use of random walks in network algorithms.

\bibitem{blondel2008fast}
V.~D. Blondel, J.-L. Guillaume, R.~Lambiotte, and E.~Lefebvre, Fast unfolding of communities in large networks. \textit{Journal of statistical mechanics: theory and experiment} \textbf{2008}(10), P10008 (2008).

\bibitem{norris1998markov}
J.~R. Norris, \textit{Markov chains}. Number~2, Cambridge university press (1998).

\bibitem{martin2012extraordinary}
C.~D. Martin and M.~A. Porter, The extraordinary svd. \textit{The American Mathematical Monthly} \textbf{119}(10), 838--851 (2012).

\bibitem{callaway2000network}
D.~S. Callaway, M.~E. Newman, S.~H. Strogatz, and D.~J. Watts, Network robustness and fragility: Percolation on random graphs. \textit{Physical review letters} \textbf{85}(25), 5468 (2000).

\bibitem{dorogovtsev2001giant}
S.~N. Dorogovtsev, J.~F.~F. Mendes, and A.~N. Samukhin, Giant strongly connected component of directed networks. \textit{Physical Review E} \textbf{64}(2), 025101 (2001).

\bibitem{cohen2021percolation}
R.~Cohen and S.~Havlin, Percolation in complex networks. \textit{Complex Media and Percolation Theory} pp. 419--431 (2021).

\bibitem{albert2002statistical}
R.~Albert and A.-L. Barab{\'a}si, Statistical mechanics of complex networks. \textit{Reviews of modern physics} \textbf{74}(1), 47 (2002).

\bibitem{rao2014protein}
V.~S. Rao, K.~Srinivas, G.~Sujini, and G.~S. Kumar, Protein-protein interaction detection: methods and analysis. \textit{International journal of proteomics} \textbf{2014}(1), 147648 (2014).

\bibitem{athanasios2017protein}
A.~Athanasios, V.~Charalampos, T.~Vasileios, and G.~Md.~Ashraf, Protein-protein interaction (ppi) network: recent advances in drug discovery. \textit{Current drug metabolism} \textbf{18}(1), 5--10 (2017).

\bibitem{burghouwt2005temporal}
G.~Burghouwt and J.~De~Wit, Temporal configurations of european airline networks. \textit{Journal of Air Transport Management} \textbf{11}(3), 185--198 (2005).

\bibitem{kosack2018functional}
S.~Kosack, M.~Coscia, E.~Smith, K.~Albrecht, A.-L. Barab{\'a}si, and R.~Hausmann, Functional structures of us state governments. \textit{Proceedings of the National Academy of Sciences} \textbf{115}(46), 11748--11753 (2018).

\bibitem{traag2019louvain}
V.~A. Traag, L.~Waltman, and N.~J. Van~Eck, From louvain to leiden: guaranteeing well-connected communities. \textit{Scientific reports} \textbf{9}(1), 5233 (2019).

\end{thebibliography}

\clearpage
\onecolumngrid
\section*{Appendix A: simulation data}
\label{sec:appendix}
\begin{table}[h!]
    \centering
    \begin{tabular}{|l|c|c|c|c|c|c|c|c|c|c|}
    \hline
    State & $N$ & $N_1$ & $C_1$ & $r(B)$ & $M_{BW}$ & $C$ & $M_{FW}$ & $\mathcal{E}$ & $N_2$ & $C_2$ \\
    \hline
    Alabama & 1127 & 1108 & 0.0169 & 51 & (410, 51) & (51, 51) & (51, 70) & 5.64 E-14 & 531 & 0.529 \\
    Alaska & 520 & 517 & 0.00577 & 23 & (136, 23) & (23, 23) & (23, 25) & 2.27 E-14 & 184 & 0.646 \\
    Arizona & 349 & 348 & 0.00287 & 48 & (48, 48) & (48, 48) & (48, 64) & 1.10 E-14 & 160 & 0.542 \\
    Arkansas & 518 & 517 & 0.00193 & 79 & (91, 79) & (79, 79) & (79, 84) & 1.99 E-14 & 254 & 0.510 \\
    California & 3585 & 3555 & 0.00837 & 349 & (860, 349) & (349, 349) & (349, 481) & 1.32 E-13 & 1690 & 0.529 \\
    Colorado & 864 & 857 & 0.00810 & 122 & (165, 122) & (122, 122) & (122, 140) & 1.66 E-14 & 427 & 0.506 \\
    Connecticut & 1377 & 1371 & 0.00436 & 157 & (289, 157) & (157, 157) & (157, 344) & 4.15 E-14 & 790 & 0.426 \\
    Delaware & 419 & 419 & 0 & 20 & (117, 20) & (20, 20) & (20, 20) & 2.08 E-14 & 157 & 0.625 \\
    Florida & 2016 & 2005 & 0.00546 & 122 & (993, 122) & (122, 122) & (122, 131) & 6.19 E-14 & 1246 & 0.382 \\
    Georgia & 1010 & 1002 & 0.00792 & 120 & (197, 120) & (120, 120) & (120, 123) & 1.50 E-14 & 440 & 0.564 \\
    Hawaii & 162 & 161 & 0.00617 & 36 & (51, 36) & (36, 36) & (36, 38) & 7.97 E-15 & 125 & 0.228 \\
    Idaho & 932 & 929 & 0.00322 & 148 & (153, 148) & (148, 148) & (148, 177) & 1.18 E-14 & 478 & 0.487 \\
    Illinois & 1017 & 1014 & 0.00295 & 99 & (172, 99) & (99, 99) & (99, 118) & 2.61 E-14 & 389 & 0.618 \\
    Indiana & 1060 & 987 & 0.0689 & 77 & (363, 77) & (77, 77) & (77, 101) & 2.41 E-14 & 541 & 0.490 \\
    Iowa & 453 & 448 & 0.0110 & 30 & (108, 30) & (30, 30) & (30, 41) & 7.89 E-15 & 179 & 0.605 \\
    Kansas & 541 & 540 & 0.00185 & 60 & (78, 60) & (60, 60) & (60, 86) & 1.61 E-14 & 224 & 0.586 \\
    Kentucky & 961 & 960 & 0.00104 & 67 & (349, 67) & (67, 67) & (67, 69) & 4.71 E-14 & 485 & 0.495 \\
    Louisiana & 405 & 395 & 0.0247 & 56 & (104, 56) & (56, 56) & (56, 66) & 1.30 E-14 & 226 & 0.442 \\
    Maine & 500 & 493 & 0.0140 & 26 & (103, 26) & (26, 26) & (26, 39) & 1.27 E-14 & 168 & 0.664 \\
    Maryland & 517 & 511 & 0.0116 & 45 & (120, 45) & (45, 45) & (45, 47) & 1.92 E-14 & 212 & 0.590 \\
    Massachusetts & 2833 & 2816 & 0.00600 & 334 & (899, 334) & (334, 334) & (334, 566) & 4.57 E-14 & 1799 & 0.365 \\
    Michigan & 1612 & 1609 & 0.00186 & 199 & (256, 199) & (199, 199) & (199, 220) & 3.21 E-14 & 675 & 0.581 \\
    Minnesota & 1395 & 1386 & 0.00645 & 203 & (279, 203) & (203, 203) & (203, 230) & 4.03 E-14 & 712 & 0.490 \\
    Mississippi & 354 & 354 & 0 & 47 & (73, 47) & (47, 47) & (47, 54) & 1.16 E-14 & 174 & 0.508 \\
    Missouri & 1040 & 1037 & 0.00288 & 130 & (142, 130) & (130, 130) & (130, 192) & 9.26 E-15 & 464 & 0.554 \\
    Nebraska & 4129 & 4088 & 0.00993 & 708 & (974, 708) & (708, 708) & (708, 1442) & 4.57 E-14 & 3124 & 0.243 \\
    Nevada & 3703 & 3665 & 0.0103 & 581 & (940, 581) & (581, 581) & (581, 1227) & 4.70 E-14 & 2748 & 0.258 \\
    New Hampshire$\,\,$ & 3968 & 3931 & 0.00932 & 630 & (953, 630) & (630, 630) & (630, 1295) & 5.88 E-14 & 2878 & 0.275 \\
    New Jersey & 1447 & 1441 & 0.00415 & 136 & (235, 136) & (136, 136) & (136, 142) & 2.07 E-14 & 513 & 0.645 \\
    New Mexico & 363 & 362 & 0.00275 & 46 & (58, 46) & (46, 46) & (46, 67) & 1.70 E-14 & 171 & 0.529 \\
    New York & 3522 & 3500 & 0.00625 & 534 & (820, 534) & (534, 534) & (534, 863) & 4.77 E-14 & 2217 & 0.371 \\
    North Carolina & 914 & 914 & 0 & 94 & (147, 94) & (94, 94) & (94, 116) & 1.19 E-14 & 357 & 0.609 \\
    North Dakota & 485 & 485 & 0 & 44 & (44, 44) & (44, 44) & (44, 89) & 4.95 E-15 & 177 & 0.635 \\
    Ohio & 3281 & 3266 & 0.00458 & 288 & (757, 288) & (288, 288) & (288, 558) & 4.54 E-14 & 1657 & 0.494 \\
    Oklahoma & 1356 & 1349 & 0.00516 & 120 & (262, 120) & (120, 120) & (120, 139) & 3.13 E-14 & 545 & 0.591 \\
    Oregon & 1610 & 1603 & 0.00435 & 158 & (277, 158) & (158, 158) & (158, 242) & 2.91 E-14 & 663 & 0.503 \\
    Pennsylvania & 2248 & 2244 & 0.00178 & 214 & (451, 214) & (214, 214) & (214, 270) & 3.22 E-14 & 1021 & 0.546 \\
    Rhode Island & 729 & 727 & 0.00275 & 70 & (96, 70) & (70, 70) & (70, 95) & 9.36 E-15 & 305 & 0.582 \\
    South Carolina & 766 & 755 & 0.0144 & 71 & (140, 71) & (71, 71) & (71, 81) & 1.64 E-14 & 370 & 0.516 \\
    South Dakota & 609 & 607 & 0.00328 & 52 & (84, 52) & (52, 52) & (52, 75) & 1.27 E-14 & 238 & 0.610 \\
    Tennessee & 2130 & 2121 & 0.00423 & 112 & (519, 112) & (112, 112) & (112, 153) & 3.34 E-14 & 1100 & 0.484 \\
    Texas & 2364 & 2356 & 0.00338 & 251 & (543, 251) & (251, 251) & (251, 319) & 4.55 E-14 & 1206 & 0.489 \\
    Utah & 4317 & 4281 & 0.00834 & 717 & (1005, 717) & (717, 717) & (717, 1447) & 4.78 E-14 & 3252 & 0.245 \\
    Vermont & 3148 & 3128 & 0.00637 & 308 & (749, 308) & (308, 308) & (308, 616) & 4.75 E-14 & 1974 & 0.371 \\
    Virginia & 2366 & 2343 & 0.00973 & 157 & (770, 157) & (157, 157) & (157, 193) & 5.21 E-14 & 1297 & 0.451 \\
    Washington & 2319 & 2304 & 0.00648 & 203 & (615, 203) & (203, 203) & (203, 270) & 4.56 E-14 & 1251 & 0.461 \\
    West Virginia & 1093 & 1093 & 0 & 94 & (135, 94) & (94, 94) & (94, 98) & 1.29 E-14 & 382 & 0.566 \\
    Wisconsin & 1552 & 1540 & 0.00773 & 230 & (316, 230) & (230, 230) & (230, 250) & 1.62 E-14 & 689 & 0.558 \\
    Wyoming & 1035 & 1028 & 0.00676 & 90 & (134, 90) & (90, 90) & (90, 91) & 2.10 E-14 & 379 & 0.633 \\
    \hline
    \end{tabular}
    \caption{Results for the website networks of the 49 of the 50 states. The network of Montana was corrupted and it's not included in the simulations. The symbols represent: $N$: number of nodes of the original network, $N_1$: number of nodes after the first compression steps, $C_1$: compression factor for the first compression steps, $r(B)$: rank of the $B$ matrix (see \cref{eq:decomposition}), $M_{BW}$: shape of the backward flow matrix, $C$: shape of the core compression matrix, $M_{FW}$: shape of the forward flow matrix, $\mathcal{E}$: compression error for the $B$ matrix, $N_2$:  number of nodes after the second compression steps, $C_1$: compression factor for the second compression steps.}\label{tab:data}
\end{table}

\end{document}